# A Novel Image Encryption using an Integration Technique of Blocks Rotation based on the Magic cube and the AES Algorithm

Ahmed Bashir Abugharsa[1], Abd Samad Bin Hasan Basari[2] and Hamida Almangush[3]

[1] Centre of Advanced Computing Technology,
Faculty of Information and Communication Technology, Universiti Teknikal Malaysia Melaka
Hang Tuah Jaya, 76100 Durian Tunggal, Melaka, MALAYSIA

[2] Centre of Advanced Computing Technology,
Faculty of Information and Communication Technology, Universiti Teknikal Malaysia Melaka
Hang Tuah Jaya, 76100 Durian Tunggal, Melaka, MALAYSIA

[3] Centre of Advanced Computing Technology,
Faculty of Information and Communication Technology, Universiti Teknikal Malaysia Melaka
Hang Tuah Jaya, 76100 Durian Tunggal, Melaka, MALAYSIA

**Abstract**
In recent years, several encryption algorithms have been proposed to protect digital images from cryptographic attacks. These encryption algorithms typically use a relatively small key space and therefore, provide safe, especially if they are of a dimension. In this paper proposes an encryption algorithm for a new image protection scheme based on the rotation of the faces of a Magic Cube. The original image is divided into six sub-images and these sub-images are divided amongst a number of blocks and attached to the faces of a Magic Cube. The faces are then scrambled using rotation of the Magic Cube. Then the rotated image is fed to the AES algorithm which is applied to the pixels of the image to encrypt the scrambled image. Finally, experimental results and security analysis show that the proposed image encryption scheme not only encrypts the picture to achieve perfect hiding, but the algorithm can also withstand exhaustive, statistical and differential attacks.

## 1. Introduction

The security of images is of particular interest in this paper. Traditional data encryption algorithms such as the private key encryption standard (DES), public key standards such as Rivest Shamir Adleman (RSA), and the family of elliptic-curve-based encryption (ECC), as well as the international data encryption algorithm (IDEA), may not be suitable for image encryption, especially for real-time applications[1]. In recent years, a number of encryption algorithms have been proposed to protect images. These encryption algorithms can be classified into several categories such as value transformation[2-5], pixel position permutation [6-9], and chaotic systems [10-13].

In the first group, Liu et al. [2] proposed an image encryption algorithm based on an iterative random phase encoding in gyrator transform domains. Two-dimensional chaotic mapping is used to create much random data for iterative random stage encoding. In [3], a colour image encryption method using a discrete fractional random transform (DFRNT) and the Arnold transform (AT) in the intensity-hue-saturation (IHS) colour space has been suggested. Each colour space component is then encrypted separately by different approaches. In [4], an image encryption algorithm based on the Arnold transform and the gyrator transform has been presented. The amplitude and stage of the gyrator transform are divided into a number of sub-images, which are shuffled using the Arnold transform. The parameters of the gyrator transforms and the separation algorithm provide the key for the encryption process.

Tao et al. [5] proposed an image encryption algorithm based on the fractional Fourier transform (FRFT) which can be applied to double or more image encryptions. The encrypted image is achieved by the summation of different orders of inverse discrete fractional Fourier transforms (IDFRFT) of the interpolated sub-images. The complete transform orders of the employed FRFT are used as the secret keys for the decryption of each sub-image.

In the second group, Zunino [6] used Peano-Hilbert curves to provide pixel position permutations (transformation) to destroy the spatial autocorrelation of an image. Zhang and Liu [7] proposed an image encryption algorithm based on a permutation-diffusion construction and a skew tent map system. In their proposed algorithm, the P-box is chosen as the size of the plain image, which totally scrambles the pixels. To enhance the security, the key stream in the diffusion step depends on both the key and the plain image. Zhao and Chen [8] proposed to used ergodic matrixes for the shuffling and encryption of images. The authors analyzed the isomorphism relationship between ergodic matrices and permutations. Zhu et al. [9] proposed an innovative permutation method to confuse and diffuse the grey-scale image at the bit level, which changes the position of each pixel and changes its value. This algorithm also utilizes the Arnold cat map to permute the bits and a logistic map to additionally encrypt the permutated image.

In the third category, Huang and Nien [10] proposed a new pixel shuffling scheme for colour image encryption which used chaotic sequences created by chaotic systems as encryption codes. In [11], a two-dimensional chaotic cat map was generalized to three-dimensions which was then utilized to design a rapid and secure symmetric image encryption algorithm. This algorithm uses the 3D cat map to scramble the locations and the values of the image pixels. Wang et al. [12] proposed an image encryption algorithm based on a simple perception and used a high-dimensional chaotic system in order to produce three sets of pseudorandom sequences. The weight of each neuron of the perception is created in addition to a set of input signals, by use of a nonlinear strategy. Recently, a new image encryption algorithm combining permutation and diffusion has been proposed by Wang et al. [13]. The original image is divided into blocks and a spatiotemporal chaotic system is then employed to create the pseudorandom sequences that are used for diffusing and scrambling these blocks.

This paper proposes an encryption algorithm of an image based on the rotation of a Magic Cube. The original image is divided into six sub-images and these sub-images are divided amongst a number of blocks and attached to the faces of a Magic Cube. The faces are then scrambled using rotation of the Magic Cube. Then the rotated image is fed to the AES algorithm which is applied to the pixels of the image to encrypt the scrambled image. Experimental results and security analysis show that the proposed image encryption scheme not only encrypts the picture to achieve perfect hiding, but the algorithm can also withstand exhaustive, statistical and differential attacks.

## 2. The Proposed Technique

### 2.1 Description of the Rotation Algorithm

The rotation algorithm is presented below. It creates a rotation table that will be used to build a newly rotated image based on the idea of the magic cube. The rotation technique works as follows:
- Load the original image and resize the image to a size of M * N so that we can partition or divide the resized image into six sub-images of the same size and with no overlap.
- The sub-images have the size (M/3) * (N/2). Mark the six faces as Up (U), Front (F), Right (R), Left (L), Down (D) and Back (B).
- Load the six sub-images and divide into a number of blocks with the same number of pixels. The image is decomposed into blocks, each one containing a specific number of pixels.
- Combine the hash function and secure key to build a rotation table of encryption that will be used to rotate the rows and columns of faces of the magic cube. The secret key and hash function of this approach are used to play a main role in building the rotated table, which will be used to generate the rotated image. The rotation process refers to the operation of dividing and rotating an arrangement of the original image.
- The main idea is that an image can be encrypted by rotating the rows and the columns of the magic cube faces (sub-images) and not to change the positions of the blocks. By rotating the rows a number of times depending on the rotation table, and then the same number of times for the columns for an arrangement of blocks, the image can be scrambled.
- For better encryption, the block size should be small, because in that way fewer pixels will be similar to their neighbours otherwise for an image with a high resolution, the content of such an image may be predicted by an unauthorized user who can thus guess the image.
- With a small block size, the correlation will be decreased and thus it becomes difficult to predict the value of any given pixel from the values of its neighbours.
- The clear information present in an image is due to the relationship (correlation) among the image elements. This perceivable information can be reduced by decreasing the correlation among the image pixels using rotation or another technique. In other words, the correlation between the blocks of the image is decreased so as to provide a good level of encryption of the image.

- At the receiver side, the original image can be retrieved by an inverse of the Rotation of the blocks. Mapping of the six sub-images on the magic cube faces is shown in Fig. 1 and a general block diagram of the rotating method is shown in Fig. 2.

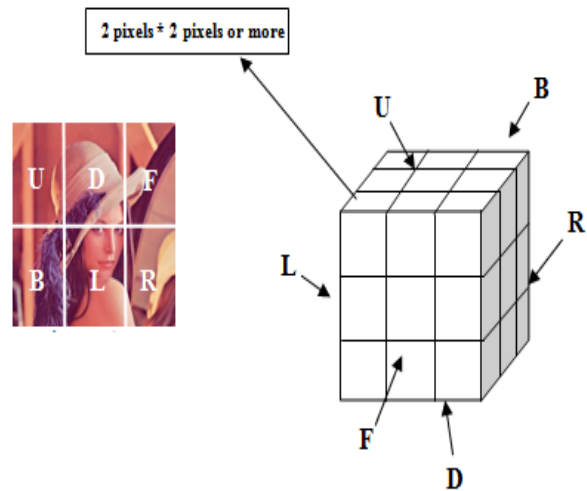

Fig. 1 Mapping of the six sub-images on the magic cube faces

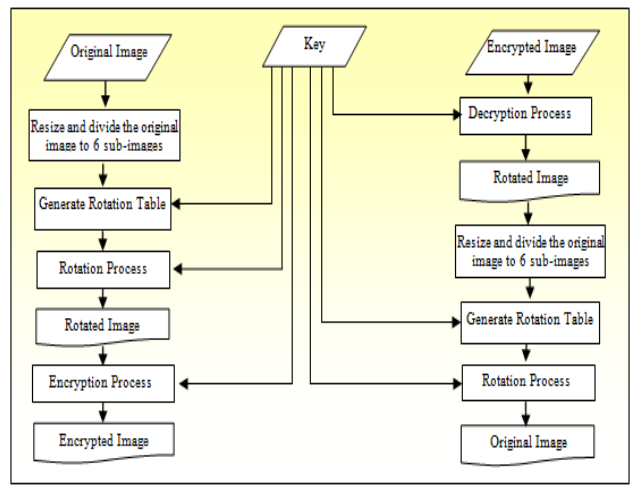

Fig. 2 Diagram of the rotation algorithm

The rotation algorithm is presented below. It creates a rotation table that will be utilized to build a newly encrypted image.

ALGORITHM CREATE_ROTATION_TABLE
1: Load Original Image
2: Input SecureKey
3: Divide the Original Image into 6 sub-images
4: Calculate Width and Height of the sub-Images
5:
    5.1: N_Horizontal = Width /3   (each block contain 3 pixels * 3 pixels)
    5.2: N_Column = Height /3    (each block contain 3 pixels * 3 pixels)

6:
    6.1: N_Column_Rotation Table (Index Of Columns in Rotation Table) = 128
    6.2: If (N_Horizontal ≥ N_Column) then
        N_Horizontal_Rotation Table(Index Of Rows in Rotation Table) = N_Horizontal
    Else
        N_Column_Rotation Table (Index Of Rows in Rotation Table) = N_Column
7:
  For I = 0 to N_Column_Rotation Table -1
    For J = 0 to N_Horizontal_Rotation Table -1
        Position Value = Hash Function (Index(I),Index(J), Secure Key)
        Position Value to Assign location I and J in the Rotation Table
    Next J
  Next I
End Create_Rotation_Table
8: Output: Rotation table

ALGORITHM CREATE_ROTATION_IMAGE (Encrypt)

1: Load Original Image
2: Input SecureKey
3: Divide the Original Image into 6 sub-images (Faces of the magic cube)
4: Calculate Image Width and Height of sub-images
5:
    5.1: N_Horizontal = Width /3
    5.2: N_Column = Height /3

6: Divide each sub-image into blocks (N_Horizontal*N_Column) (each block contain 3 pixels * 3 pixels).
7: L_Key = Length (SecureKey)
8:
  For J = 0 to L_Key-1
    8.1: (Rotation of The Rows Of Images that are attached to the faces of the magic cube F, U, B, D) ndexOfColumnsInRotationTable= Int (SecureKey( J ))
      For I = 0 to N_Horizontal -1
        NumberOfRotation = RotationTable(I, IndexOfColumnsInRotationTable )
        Rotate all the rows I in all the images F, U, B, D of the magic cube (NumberOfRotation).
      Next I
    8.2: (Rotation of The Columns Of Images that are attached to the faces of the magic cube F, R, B, L)
      IndexOfColumnsInRotationTable= Int (SecureKey( J ))
      For I = 0 to N_Column -1
        NumberOfRotation = RotationTable(I, IndexOfColumnsInRotationTable )
        Rotate all the columns I in all the images F, R, B, L of the magic cube (NumberOfRotation).
      Next I
  Next J
End Create_Rotation_Image
9: Output: Rotation Image (Image Encryption)

## 2.2 Description of Integration Technique

The block-based rotation algorithm is based on the integration of image rotation followed by the AES algorithm. The rotation algorithm and the AES algorithm use the original image to generate three encrypted images; (a) a ciphered image using the AES algorithm, (b) a rotation image using a rotation process and (c) a rotation image encrypted using the AES algorithm.

The correlation and entropy of the three images are calculated and evaluated against each other. This technique aims at producing a good security level for the encrypted images by decreasing the correlation among the image pixels and increasing its entropy value. Image measurements (correlation, entropy and differential analysis) will be carried out on the original image and the encrypted images with and without the rotation algorithm and the results will then be analyzed. The overview of the integration technique is presented in Fig. 3.

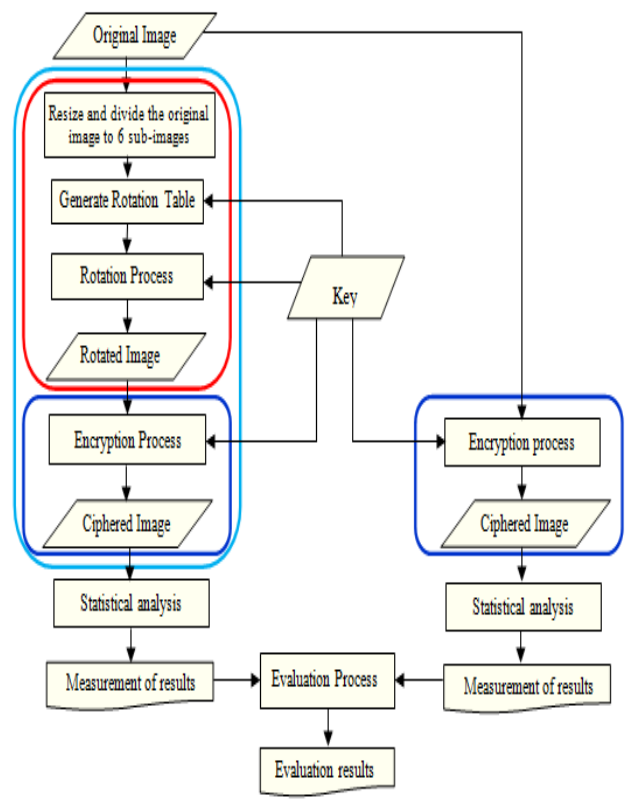

Fig. 3. Diagram of the proposed technique

— Rotation algorithm    — AES algorithm    — Proposed technique

# 3. Experimental Details and Results

A good quality encryption algorithm should be strong against all types of attack, including statistical and brute force attacks. Some experiments are given in this section to demonstrate the efficiency of the proposed technique. In this section, the proposed technique is applied on an image that has 300 * 300 pixels and four selected different cases are analyzed in detail to test the performance of the proposed technique. The number of blocks and the block sizes in each case are shown in Table 1.

Table 1 Different cases of number of blocks and the number of pixels

| Case number | Number of blocks | Block size |
|---|---|---|
| 1 | 150 * 150 | 2 Pixels * 2 Pixels |
| 2 | 100 * 100 | 3 Pixels * 3 Pixels |
| 3 | 60 * 60 | 5 Pixels * 5 Pixels |
| 4 | 50 * 50 | 6 Pixels * 6 Pixels |

The rotation algorithm and the AES algorithm are used on the plain image to generate three encrypted images (a) a ciphered image using the AES algorithm, (b) a rotation image using the rotation process and (c) a rotation image encrypted using the AES algorithm. The correlation and entropy of the three images are calculated and evaluated.

## 2.3 Statistical Analysis

In order to resist statistical attacks, the encrypted images should possess certain random properties. To prove the robustness of the proposed algorithm, a statistical analysis has been performed by calculating the histograms, the entropy, the correlations and differential analysis for the plain image and the encrypted image. Different images have been tested, and it has been determined that the intensity values are good.

### 2.3.1 Histogram Analysis

An image histogram is a commonly used method of analysis in image processing and data mining applications. The advantage of a histogram is that it shows the shape of the distribution for a large set of data. Thus, an image histogram illustrates how the pixels in an image are distributed by graphing the number of pixels at each colour intensity level. It is important to ensure that the encrypted and original images do not have any statistical similarities. The histogram analysis clarifies how pixels in an image are distributed by plotting the number of pixels at each intensity level.

The experimental results of the plain image and its corresponding cipher image and their histograms are shown in Fig. 4. The histogram of each plain image illustrates how the pixels are distributed by graphing the number of pixels at every grey level [14]. It is clear that the histogram of the encrypted image is nearly uniformly distributed, and significantly different from the respective histograms of the original images.

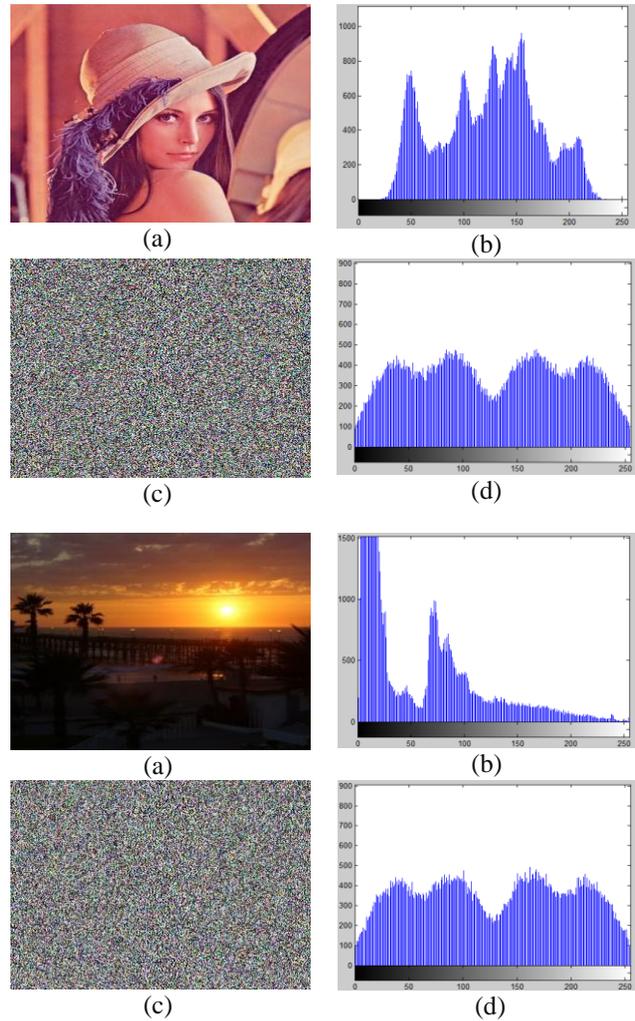

Fig. 4 : (a) Original Image (b) Histogram of Original Image (c) Encrypted Image (d) Histogram of Encrypted Image

### 2.3.2 Correlation of two adjacent pixels

A correlation is a statistical measure of security that expresses a degree of relationship between two adjacent pixels in an image or a degree of association between two adjacent pixels in an image. The aim of correlation measures is to keep the amount of redundant information available in the encrypted image as low as possible[15, 16].

In general, if the correlation coefficient equals zero or is very near to zero, then the original image and its encrypted version are totally different. It can be inferred that the encrypted image has no features and is highly independent of the original image. If the correlation coefficient is equal to -1, that means the encrypted image is a negative of the original image.

In the experiments results, 2000 pairs of two adjacent pixels are randomly selected. Fig. 5 shows the distribution of two adjacent pixels in the original image and the encrypted-image. There is very good correlation between adjacent pixels in the image data [17, 18], while there is only a small correlation between adjacent pixels in the encrypted image. Equation (1) is used to study the correlation between two adjacent pixels in the horizontal, vertical, diagonal and anti-diagonal orientations:

$$C_r = \frac{N\sum_{j=1}^{N}(X_j \times Y_j) - \sum_{j=1}^{N}X_j \times \sum_{j=1}^{N}Y_j}{\sqrt{\left(N\sum_{j=1}^{N}X_j^2 - \left(\sum_{j=1}^{N}X_j\right)^2\right) \times \left(N\sum_{j=1}^{N}Y_j^2 - \left(\sum_{j=1}^{N}Y_j\right)^2\right)}} \quad (1)$$

where *x* and *y* are the intensity values of two neighbouring pixels in the image and *N* is the number of adjacent pixels selected from the image to calculate the correlation. Results for the correlation coefficients of two adjacent pixels are shown in Tables 2, 3, 4 and 5.

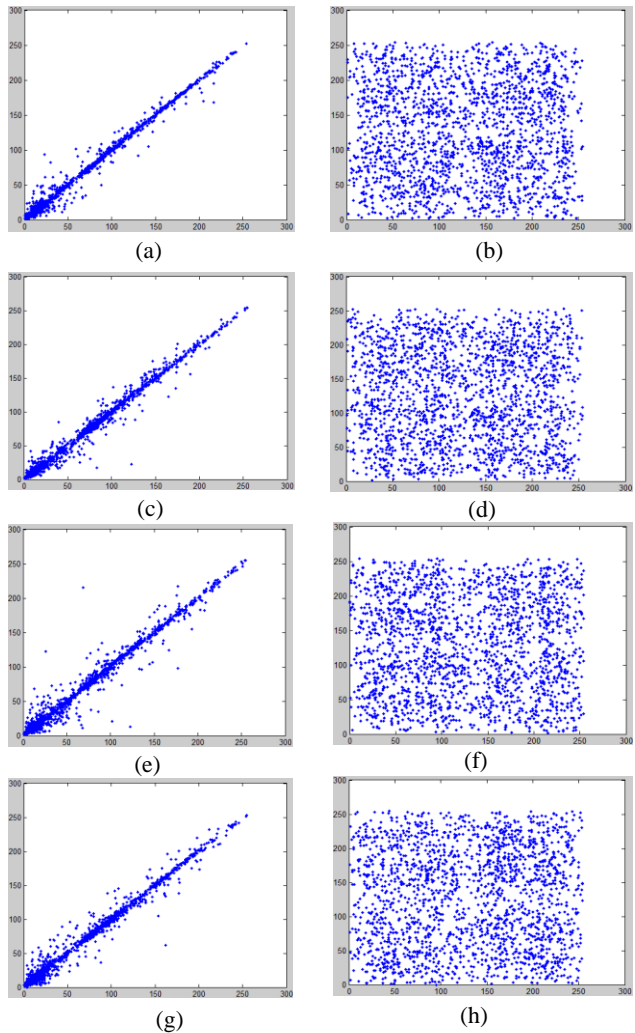

Fig.5: Correlation of two adjacent pixels: (a) distribution of two horizontally adjacent pixels in the original image, (b) distribution of two horizontally adjacent pixels in the encrypted image (i.e., cipher image); (c) distribution of two vertically adjacent pixels in the original image, (d) distribution of two vertically adjacent pixels in the encrypted image. (e) distribution of two diagonally adjacent pixels in the original image. (f) distribution of two diagonally adjacent pixels in the encrypted image. (g) distribution of two anti-diagonally adjacent pixels in the original image. (h) Distribution of two anti-diagonally adjacent pixels in the encrypted image.

### 2.3.3 Information Entropy

Information theory is the mathematical theory of data communication and storage founded in 1949 by Shannon [19]. Information entropy is defined to express the degree of uncertainties in the system. It is well known that the entropy $H(m)$ of a message source *m* can be calculated as:

$$H(m) = \sum_{i=0}^{2N-1} P(m) \log_2 \frac{1}{P(m_i)} \quad (2)$$

where $P(m_i)$ represents the probability of symbol $m_i$ and the entropy is expressed in bits. Let us suppose that the source emits $2^8$ symbols with equal probability, i.e., $m = \{m_1, m_2, ..., m_{2^8}\}$. Truly random source entropy is equal to 8. Actually, given that a practical information source seldom generates random messages, in general its entropy value is smaller than the ideal. However, when the messages are encrypted, their entropy should ideally be 8. If the output of such a cipher emits symbols with an entropy of less than 8, there exists a certain degree of predictability which threatens its security. Results for the entropy are shown in Tables 2, 3, 4 and 5.

**Case 1**: The image is divided into 6 pixels * 6 pixels in each block. Figure 6 shows the image cases:

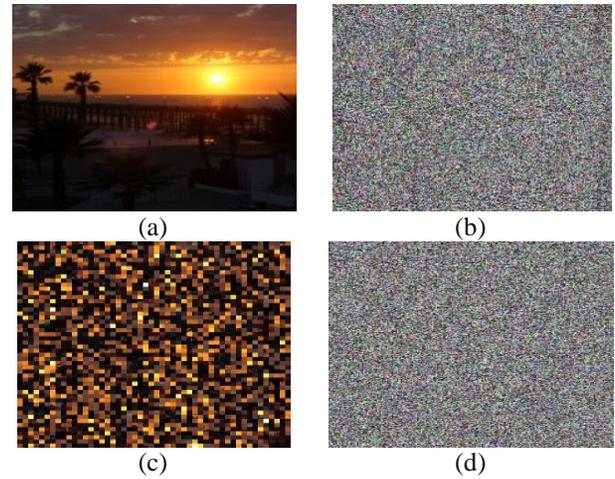

Fig. 6. (a) Original image. (b) Encrypted image using AES (c) Rotation image. (d) Encrypted image using integration technique

Table 2. Correlation of Two Pixels and Entropy value

| | Correlation Analysis | | | | Entropy value |
|---|---|---|---|---|---|
| Image | adjacent pixels | | | | |
| | Horizontal | Vertical | Diagonal | Anti-Diagonal | |
| A | 0.9951 | 0.99355 | 0.9917 | 0.9914 | 7.0443 |
| B | -0.04444 | -0.0476 | -0.0208 | -0.0317 | 7.9256 |
| C | 0.7540 | 0.7790 | 0.6120 | 0.6389 | 7.1338 |
| D | -0.0456 | -0.0478 | -0.0253 | -0.0319 | 7.9417 |

**Case 2**: The image is divided into 5 pixels * 5 pixels in each block. Fig. 7 shows the image cases:

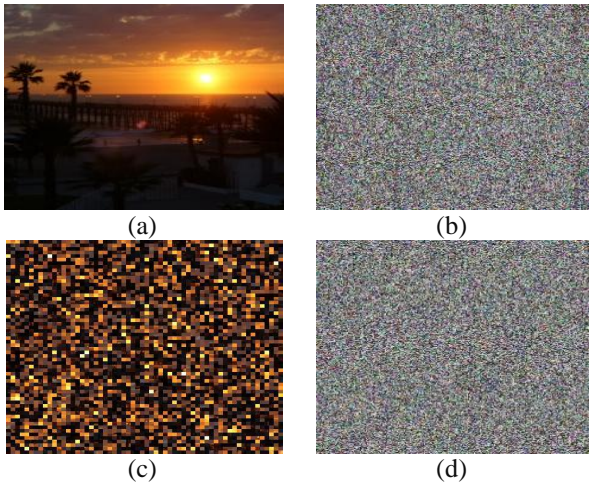

(a)  (b)
(c)  (d)

Figure 7. (a) Original image. (b) Encrypted image using AES (c) Rotation image. (d) Encrypted image using integration technique.

Table 3. Correlation of Two Pixels and Entropy value

| Image | Correlation Analysis ||||Entropy value |
|---|---|---|---|---|---|
| | adjacent pixels ||||  |
| | Horizontal | Vertical | Diagonal | Anti-Diagonal | |
| A | 0.9951 | 0.99355 | 0.9917 | 0.9914 | 7.0443 |
| B | -0.0444 | -0.0476 | -0.0208 | -0.0317 | 7.9256 |
| C | 0.7420 | 0.7577 | 0.5920 | 0.5998 | 7.1538 |
| D | -0.0468 | -0.0488 | -0.0353 | -0.0386 | 7.9438 |

**Case 3**: The image is divided into 3 pixels * 3 pixels in each block. Figure 8 shows the image cases:

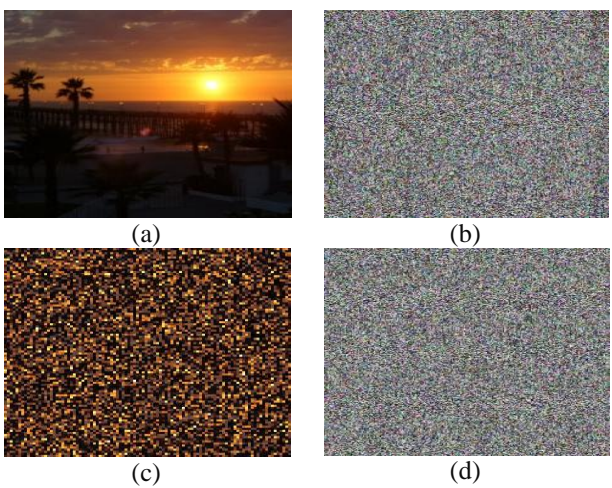

(a)  (b)
(c)  (d)

Fig. 8. (a) Original image. (b) Encrypted image using AES (c) Rotation image. (d) Encrypted image using integration technique.

Table 4. Correlation of Two Pixels and Entropy value

| Image | Correlation Analysis ||||Entropy value |
|---|---|---|---|---|---|
| | adjacent pixels ||||  |
| | Horizontal | Vertical | Diagonal | Anti-Diagonal | |
| A | 0.9951 | 0.99355 | 0.9917 | 0.9914 | 7.0443 |
| B | -0.04444 | -0.0476 | -0.0208 | -0.0317 | 7.9256 |
| C | 0.5920 | 0.5887 | 0.3820 | 0.3898 | 7.1766 |
| D | -0.0449 | -0.0520 | -0.0433 | -0.0388 | 7.9448 |

**Case 4**: The image is divided into 2 pixels * 2 pixels in each block. Figure 9 shows the image cases:

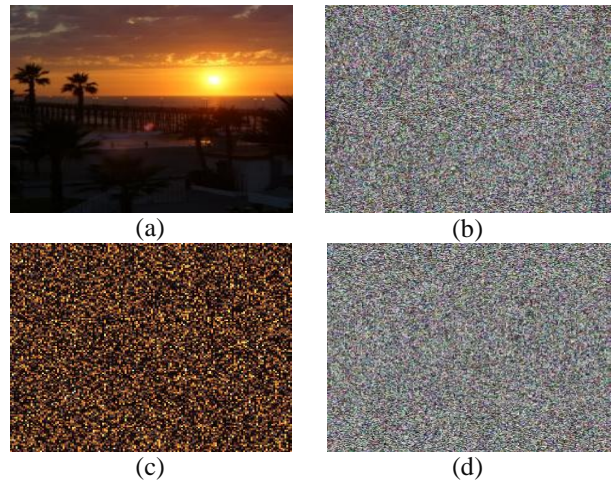

(a)  (b)
(c)  (d)

Fig. 9. (a) Original image. (b) Encrypted image using AES (c) Rotation image. (d) Encrypted image using integration technique.

Table 5. Correlation of Two Pixels and Entropy value

| Image | Correlation Analysis ||||Entropy value |
|---|---|---|---|---|---|
| | adjacent pixels ||||  |
| | Horizontal | Vertical | Diagonal | Anti-Diagonal | |
| A | 0.9951 | 0.99355 | 0.9917 | 0.9914 | 7.0443 |
| B | -0.04444 | -0.0476 | -0.0208 | -0.0317 | 7.9256 |
| C | 0.5812 | 0.5773 | 0.3820 | 0.3898 | 7.1766 |
| D | -0.0449 | -0.0520 | -0.0433 | -0.0388 | 7.9448 |

## 4.Conclusion

In this paper, a new image encryption algorithm is proposed. This algorithm is based on the theory of the Magic cube to shuffle the image blocks. To confuse the relationship between the plain image and the encrypted image, the rotated image is fed into an AES algorithm which is applied to each pixel of the image to encrypt the image even further. Experimental tests have been carried out utilising detailed numerical analysis which shows the strength of the proposed algorithm against several types of attack such as statistical and differential attacks. The proposed technique presented an inverse relationship existing between the number of blocks and correlation. There exists a direct relationship between the number of blocks and entropy. This technique is expected to show good performance, uniform distribution in a histogram, a low correlation and high entropy. Moreover, performance assessment tests demonstrate that the proposed image encryption algorithm is highly secure. It is also able to encrypt large data sets efficiently. The proposed method is expected to be useful for real time image encryption and transmission applications.

**Acknowledgments**

This paper is part of PhD work in the Faculty of Information and Communication Technology, Universiti Teknikal Malaysia Melaka (UTeM).

Ahmed Bashir Abugharsa received BSc in Computer Science from Misurata University in Misurata, Libya, MSc from Universiti Tun Abd Razak, Faculty of Information Technology in January 2011 in Kuala Lumpur, Malaysia and currently enrolled in the PhD program in Computer Science in the Universiti Teknikal Malaysia Melaka (UTeM) in Malaka, Malaysia.

Dr. ABD. SAMAD BIN HASAN BASARI received BSc in Mathematics from Universiti Kebangsaan Malaysia in 1998, Master in IT-Education from Universiti Universiti Teknologi Malaysia in 2002, PhD in ICT from Universiti Teknikal Malaysia Melaka in 2009 and currently SENIOR Department of Industrial Computing Faculty of Information and Communication Technology, Universiti Teknikal Malaysia Melaka (UTeM).

Hamida Mohamed Almangush received BSc in Computer Science from Misurata University in Misurata, Libya, MSc from Universiti Tun Abd Razak, Faculty of Information Technology in January 2011 in Kuala Lumpur, Malaysia and currently enrolled in the PhD program in Computer Science in the Universiti Teknikal Malaysia Melaka (UTeM) in Malaka, Malaysia.